\documentclass[journal]{IEEEtran}
\usepackage{amsmath,amsfonts}
\usepackage{algorithmic}
\usepackage{algorithm}
\usepackage{array}
\usepackage[caption=false,font=normalsize,labelfont=sf,textfont=sf]{subfig}
\usepackage{textcomp}
\usepackage{stfloats}
\usepackage{url}
\usepackage{verbatim}
\usepackage{graphicx}
\usepackage{cite}
\usepackage{xcolor}
\usepackage{color}

\begin{document}

\title{Integrated Optical Electric Field Sensors: Humidity Stability Mechanisms and Packaging Scheme}

 \author{
    Xinyu Ma,
	Chijie Zhuang,
  	She Wang,
	 and Rong Zeng

 	\thanks{	
		This work was supported in part by  the National Key Research and Development Program of China under Grant 2022YFB3206801, and National Natural Science Foundation of China under Grant 52022044. \emph{(Corresponding author: Chijie Zhuang.)}
		
 		X. Ma, C. Zhuang, and R. Zeng are with the State Key Lab of Power Systems, Department of Electrical Engineering, Tsinghua University, Beijing 100084, China (e-mail: \mbox{mxy18@tsinghua.org.cn;} \mbox{chijie@tsinghua.edu.cn;} \mbox{zengrong@tsinghua.edu.cn}). 

        S. Wang is with China International Engineering Consulting Corporation, Beijing 100048, China (e-mail: \mbox{wangshe@ciecc.com.cn}).
	}
 }
\markboth{}%
{Xinyu Ma \MakeLowercase{\textit{et al.}}: Submitted to J. Phys. D: Appl. Phys.}
\maketitle
	
\begin{abstract}
Integrated optical electric field sensors (IOES) play a crucial role in electric field measurement. {This paper introduces the principles of the IOES and quantitatively evaluates the impact of humidity on measurement accuracy. 
Sensors with different levels of hydrophobicity coatings and hygroscopicity shells are fabricated and tested across the relative humidity (RH) range of 25\% to 95\%. Results reveal that humidity stability is primarily influenced by water vapor absorption through the sensor shell, which increases its conductivity. This further results in amplitude deviation and phase shift of the sensor output. To address this, an optimal humidity-stable packaging scheme is proposed, which involves using PEEK shell with room temperature vulcanized fluorinated silicone rubber coating. Compared with uncoated ceramic shell, the phase shift of the IOES reduces from 90$^\circ$ to 1$^\circ$ under a RH of 90\%.} The amplitude deviation of electric field measurement decreases from 20\% to nearly zero after a 20-hour humidity experiment conducted under RH of 90\% at 30 $^\circ$C. The proposed packaging scheme could be used to improve the humidity stability of the sensors deployed in outdoor environments, especially on ships and coastal areas.
\end{abstract}


\begin{figure*}[!t]\centering
	\includegraphics[width=\linewidth]{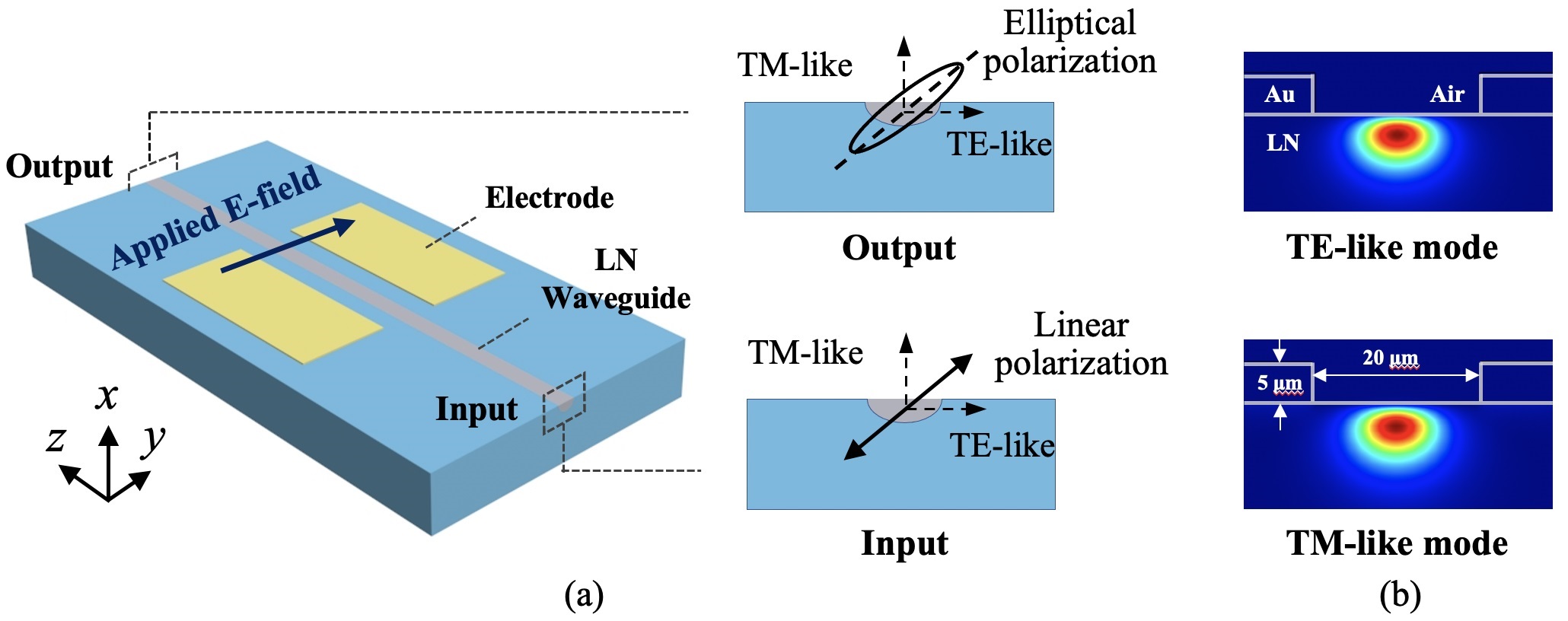}
	\caption{The schematic diagram of the IOES. (a) Principle of the IOES based on common-path interferometer. (b) Optical mode profiles of TE-like and TM-like modes.}\label{Fig1_IOES}
\end{figure*}

\section{Introduction}
With the flourishing development of the electric power industry, integrated optical electric field sensors (IOES) have gained widespread usage in electric field measurements\cite{zeng_2012sensors}. These sensors offer several advantages, including wide frequency range, compact size, and non-contact measurement capabilities. They have been applied in various scenarios, such as insulator testing \cite{zeng2008measurement}, transient overvoltage in substations \cite{wang2018transient}, very fast overvoltage in GIS \cite{zeng2009application},  \textcolor{black}{electromagnetic field around power line \cite{yuan_tim2018}, AC and DC electric field \cite{fan_tim2021} and intense electromagnetic pulse measurement \cite{shi_tim2021}, \cite{zhang_2023tim}, \cite{lee_2023tim}. }

However, the environmental stability of IOES based on electro-optic crystals, with lithium niobate (LN) being the most widely used \textcolor{black}{\cite{gutierrez_tim2008}}, has always been a concern \cite{nagata1995progress},\cite{ma2020large}. When the sensors are deployed in outdoor environments, fluctuations in the temperature and humidity can negatively impact the measurement accuracy by affecting the bias point of the sensor. A large number of researchers have conducted studies on the temperature stability of the sensors due to the physical effects exhibited by electro-optic crystals, such as pyroelectric effect and thermo-optic effect \cite{wang2019thermal}, \textcolor{black}{\cite{li_sensor2023}}.

The humidity stability of the IOES refers to its ability to maintain unaltered output under varying humidity levels, including amplitude and phase angle. In 1986, Preston, et al. found that high humidity levels could affect the device reliability, leading them to propose the hermetic fiber-tailed package for LN electro-optic devices \cite{preston1989high}. Subsequent studies primarily focused on improving sensor humidity stability through design of the hermetic packaging structure \cite{moyer1998design},\cite{johnson1996reliability}. Until 2009, Li found that even with hermetical packaging, the response characteristics of LN IOES could still be affected by the relative humidity (RH) in the environment \cite{huan2009theoretical}. In 2013, Li, et al. further demonstrated through experiments that the stability of IOES is affected by humidity. He proposed that the continuous thickening of water film on sensor { shell} surface with increasing humidity is the underlying reason that affects humidity stability \cite{li2013improvement}.

The researcher thinks that the humidity stability of the sensor depends on the hydrophobicity of its {shell}. Through the application of a hydrophobic (HP) coating, achieved by treating the {shell} surface with epoxide-resin glue, the contact angle (CA) of the {shell} surface increased from 40$^\circ$ to 85$^\circ$. Consequently, water molecules on the { shell surface} cannot form continuous water film. This modification enables the sensor to withstand higher levels of { RH}, reaching up to 85\% at 20 $^\circ$C and 50\% at 30 $^\circ$C \cite{li2013improvement}. However, the humidity range still limits the application of IOES, necessitating further research into the humidity stability.

It is also proposed that while a HP film can prevent the penetration of liquid water, it has never been effective for the water vapor transmission through HP materials \cite{fan2009experimental}. HP coatings, including those like room temperature vulcanized silicone rubber (RTV-SIR) \cite{momen2011wettability}, may contain small pores that can absorb moisture from the surrounding air \cite{fan2009experimental},\cite{wang2016moisture}. Consequently, this absorbed moisture can condense into liquid water and dissolve into the shell material \cite{shirangi2010mechanism}. The extent of water absorption increases in correlation with both { RH} levels and the porosity of the coating \cite{bond2006modeling},\cite{shook1996diffusion}, thereby impacting the electrical properties of the sensor, such as dielectric constant and conductivity \cite{christie2009electrical}.

\textcolor{black}{In our experiments, it is observed that the phase shift of the IOES increases to 90$^\circ$ under the RH of 90\%, resulting in the invalidation of IOES. With HP coatings, the phase shifts decrease to about 30$^\circ$, which still degrades the measurement accuracy of IOES. Our theoretical models, focused on water vapor thickness and sensor shell conductivity, reveal that the key factor impacting humidity stability is the sensor shell's hygroscopicity, which causes water vapor absorption and increased conductivity. It contrasts with prior belief attributing humidity stability to sensor { shell} hydrophobicity. We propose an optimal packaging scheme with a combination of low hygroscopicity in the shell material and low moisture permeability in the HP coating, which improves the phase shift of the IOES to 1$^\circ$ under the RH of 90\%. The amplitude deviation of the electric field measurement decreases from 20\% to nearly zero after 20 hours testing.}

The paper is organized as follows. Section II introduces the principles of the IOES and explores the influence of humidity on the sensor accuracy. Three sensors with different HP coatings are fabricated, and the humidity stability is evaluated. Section III studies the influence mechanisms of the sensor humidity stability by theory and modeling. In Section IV, an optimal packaging scheme is proposed and validated through a long-term testing in a high humidity environment. The proposed packaging scheme to improve the humidity stability could be extended to other sensors as well.

\section{Principles and Measurement Results}

\subsection{Principle of IOES}
The principle of the IOES is based on the Pockels effect of LN, which means the refractive index of electro-optic crystal changes linearly with the applied electric field. 
{\color{black}The schematic diagram of the IOES is shown in \mbox{Fig. \ref{Fig1_IOES}(a)}. The sensor is based on x-cut bulk LN, and light propagates along z-axis. The principle of the IOES is based on common-path interferometer \cite{zeng_2012sensors}. The input light is linear polarized by a fiber polarizer, which could be divided into the transverse electric (TE)-like mode and transverse magnetic (TM)-like mode. The former is polarized in y direction, while the latter is polarized in x direction. The optical mode profiles are shown in \mbox{Fig. \ref{Fig1_IOES}(b)}. The waveguide is fabricated by Ti-diffusion with a 7-mm-width titanium stripe. Because of the phase difference of the two modes after propagating through the same waveguide path, the output light is elliptical polarized. A fiber analyzer is used at the output, and then a photodetector converts the optical power into voltage.}

The transfer function of the IOES between the applied electric field $E_{0}$ and the sensor output $V_{\rm out}$ is

\begin{align}
V_{\rm out} = A \cdot \left[1+b \cos \left(\varphi_0 + \varphi \left( E_0 \right) \right) \right].
\label{eq_111}
\end{align}

\noindent
where $A$ is the photoelectric conversion coefficient, and $b$ is the extinction ratio.
$\varphi_0 = \frac{2\pi}{\lambda}\left(n_{\rm TE}-n_{\rm TM} \right)L$ is the bias point without applied electric field. $n_{\rm TE}$ and $n_{\rm TM}$ are the effective refractive indexes of the TE and TM-like modes. $\lambda$ is the optical wavelength, and $L$ is the length of the electrode. $\varphi(E) = -\frac{2\pi}{\lambda}\gamma_{22} n_{\rm o}^3 E_0 L$ is the phase change induced by the applied electric field $E_{0}$, where $\gamma_{22}$ is the component of the electro-optical coefficient of LN, and $n_{\rm o}$ is the ordinary index of LN \cite{wang2019thermal}.

By designing the width and length of the waveguide path, the bias point $\varphi_0$ could be set as $\pi$/2. Then Eq. (\ref{eq_111}) is converted into

\begin{align}
V_{\rm out} = A \cdot \left( 1-b \cdot \frac{2\pi}{\lambda}\gamma_{22} n_o^3 E_0 L \right).
\label{eq1}
\end{align}

{The output voltage $V_{\rm out}$ is proportional to the applied electric field $E_{0}$.}

{\subsection{Experimental setup and influence of humidity}}

\begin{figure}[!htb]\centering
	\includegraphics[width=8.5cm]{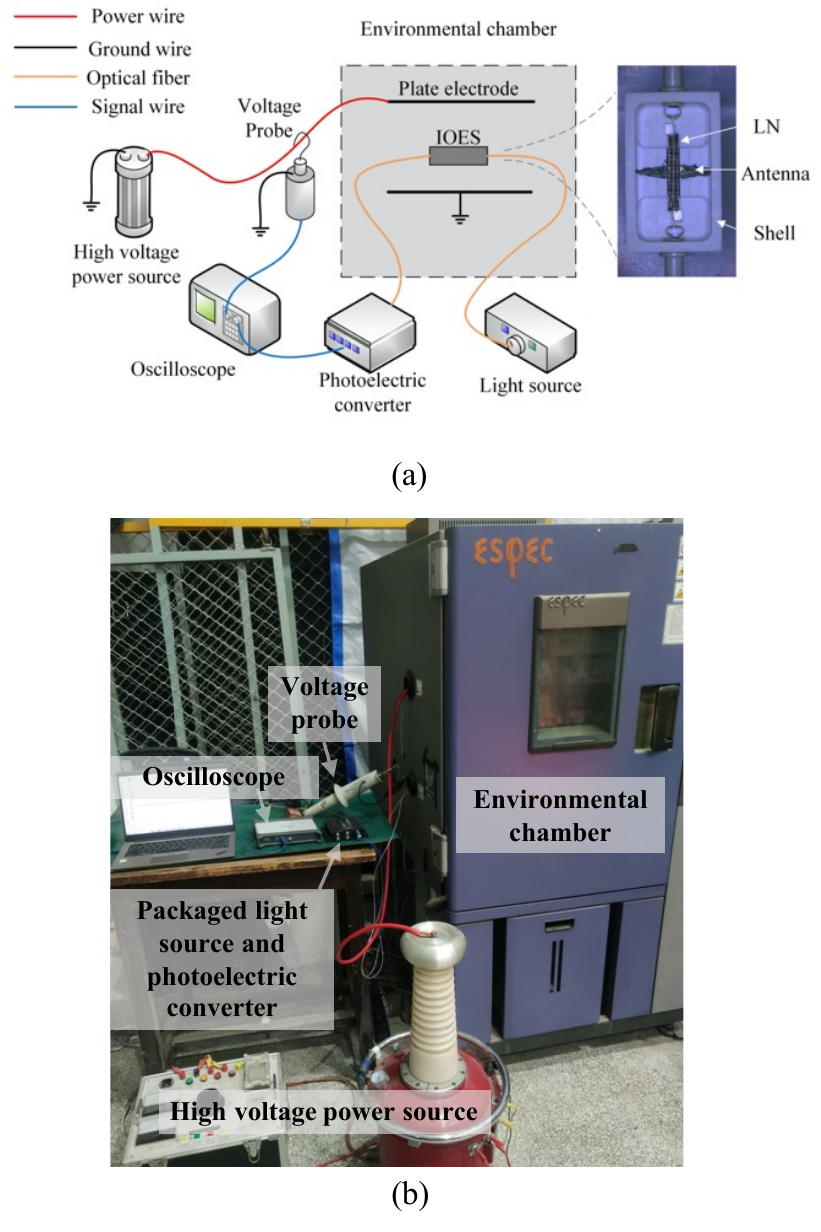}
	\caption{The humidity stability experimental setup for IOES. (a) Schematic diagram. {The inset photo shows the sensor inside. (b) Photograph.}}\label{Fig1}
\end{figure}

{The experimental setup for humidity stability is shown in a schematic diagram and a photograph in \mbox{Fig. \ref{Fig1}}(a) and (b), respectively. The inset photograph in \mbox{Fig. \ref{Fig1}}(a) shows the sensor inside, which inludes LN crystal, metal antenna and sensor shell (before coated).} The environmental chamber (Espec, model EL-02KA) has a temperature (T) range from -40 to 150 $^\circ$C with a control accuracy of $\pm$ 0.5 $^\circ$C and a RH range from 10\% to 100\%. A 50 Hz power-frequency high-voltage source is connected to a plate electrode, which is settled in the environmental chamber
A high-voltage probe is used to measure the applied voltage. The IOES is placed between the plate electrodes, and connected to the outside {light source} and photoelectric converter through polarization maintaining fibers. 
{The light source and photoelectric converter are packaged together and custom-made by Discoveryoptica Inc. The output power of light source is 250 $\mu$W and the total insertion loss of IOES is about 3 dB.}
{The utilized optical wavelength is 1310 nm due to its compatibility with low-loss fibers and easy accessibility as a light source.} The applied voltage and output of the IOES are simultaneously measured by the oscilloscope.

\begin{figure}[!t]\centering
	\includegraphics[width=8.5cm]{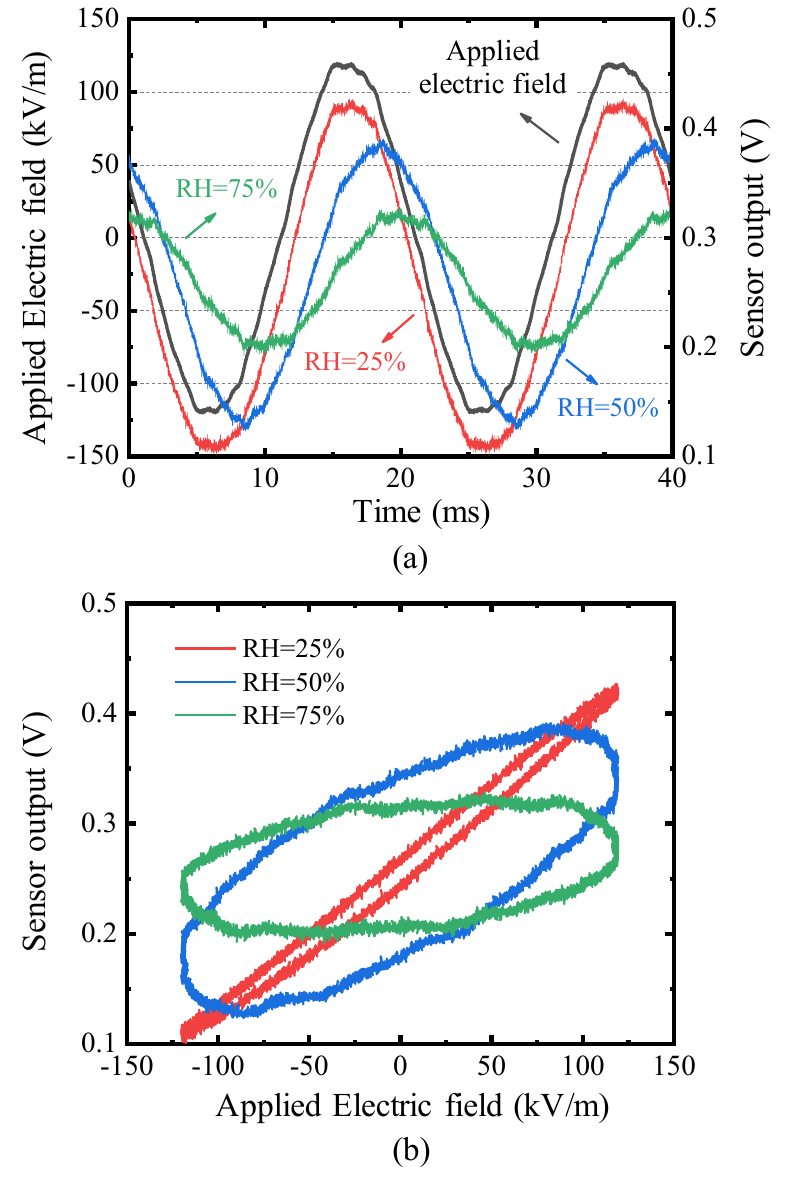}
	\caption{{ Measured} sensor outputs at 50 Hz under different RH levels. (a) Sensor outputs in the time domain. (b) Input-output characteristics.}\label{Fig2}
\end{figure}

{Measured} sensor outputs under different humidity levels are shown in \mbox{Fig. \ref{Fig2}}. With the increment of RH levels {at a rate of 5\%/h}, the phase shift becomes larger and the output voltage degrades obviously. When the RH reaches 75\%, the output voltage decreases to half of the applied electric field. {\mbox{Fig. \ref{Fig2}} (b) shows the input-output characteristics corresponding to \mbox{Fig. \ref{Fig2}} (a), also known as Lissajous curve. This shows the phase shift between applied electric field and sensor output more obviously.} Therefore, the humidity strongly influences the accuracy of IOES.

In order to quantitatively evaluate the influence of humidity, the figure of merit should be settled first. The amplitude deviation $\Delta k$ is defined as:

\begin{align}
\Delta k = \frac{E_{0}-E_{1}}{E_{0}}.
\end{align}
\noindent 
where $E_{1}$ is the amplitude of measured electric field. $E_{0}$ is obtained by the high-voltage probe and $E_{1}$ is derived by the output of the IOES refers to (1) {and (2).} {As the output signal of IOES is voltage, the  measured electric field derived by the IOES after drying is used for calibrating measured electric fields.}

The phase shift $\Delta\varphi$ is defined as:

\begin{align}
\Delta \varphi = \varphi_{0}-\varphi_{1}.
\end{align}
\noindent
where $\varphi_{0}$ and $\varphi_{1}$ are the phase angles of applied and measured electric field.

\begin{figure}[!t]\centering
	\includegraphics[width=8.5cm]{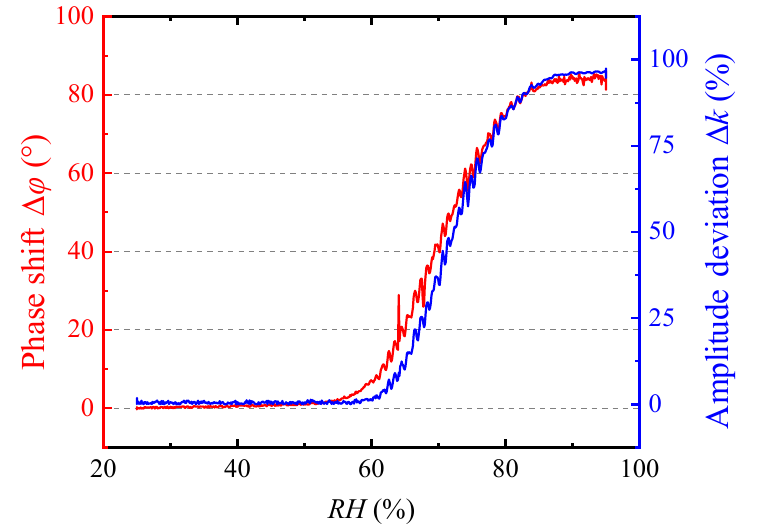}
	\caption{{Measured} variations in amplitude deviation $\Delta k$ and phase shift $\Delta\varphi$ of an IOES as a function of relative humidity at 30 $^\circ$C.}\label{Fig3}
\end{figure}

{Measured} variations in amplitude deviation $\Delta k$ and phase shift $\Delta \varphi$ of an IOES at a temperature of 30 $^\circ$C is shown in \mbox{Fig. \ref{Fig3}}. These two parameters first keep constant in a low RH level, then increase exponentially as RH increases, and finally saturate to nearly 100\% and 90$^\circ$, respectively. It is shown that the amplitude deviation $\Delta k$ and phase shift $\Delta \varphi$ change in the same trend. Since the amplitude deviation $\Delta k$ changes with different applied electric field $E_{0}$, {which could make $\Delta k$ varies in different cases (e.g., applied voltage, metal plate gap),} phase shift $\Delta \varphi$ is selected as the figure of merit to characterize the humidity stability of the sensor in the rest of the paper. Smaller phase shift suggests better humidity stability.

\subsection{{Influence} of { shell} hydrophobicity}

The influence of {shell} hydrophobicity on the humidity stability is explored with different HP coatings, providing insights for the design of the humidity-stable packaging schemes. {The coatings are applied to the outer surface of sensor shell shown in the inset photograph of \mbox{Fig. \ref{Fig1}}(a).}

The sensor coating is required to exhibit three properties: good hydrophobicity, strong adhesion to the shell, and insulation. Three HP coatings with different hydrophobicity levels were selected: RTV-SIR, room temperature vulcanized fluorinated silicone rubber (RTV-SIFR), and super-hydrophobic material (SHP) \cite{qing2020microskeleton}. The shell material of the sensor is maintained as ceramics.

Before coating the sensors, a thorough cleaning with alcohol is performed, followed by placing them in a drying oven at 80 $^\circ$C for 10 hours to prevent any potential moisture within the shells. Subsequently, the three HP coatings are coated uniformly to the surface of the ceramic shell at room temperature. The coatings are then naturally solidified for 5 hours and subsequently dried in the drying oven again at 60$^\circ$C for 40 hours. These treatments ensure stable adhesion of the HP coatings to the sensor shells.

\begin{figure}[!htb]\centering
	\includegraphics[width=8.5cm]{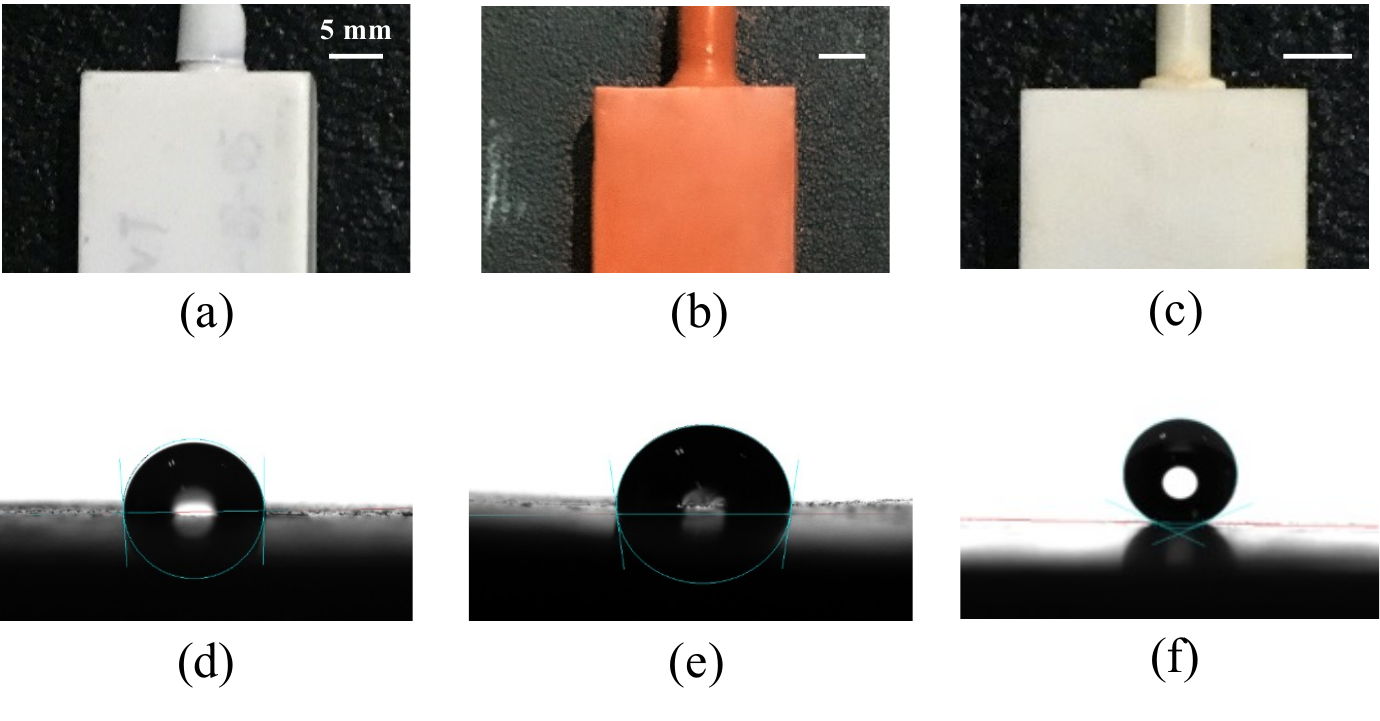}
	\caption{Sensors with different coatings (a-c) and corresponding contact angle measurement results (d-e). Sensors coated with (a) RTV-SIR, (b) RTV-SIFR and (c) SHP. {The sensors are of the same size. The scale bars indicate 5 mm in length.} Contact angles of (d) RTV-SIR, (e) RTV-SIFR and (f) SHP.}\label{Fig4}
\end{figure}

The hydrophobicity is assessed using contact angle (CA) and sliding angle measurements, which are determined by a contact angle meter (Dataphysics OCA15 pro) with 4 ${\rm \mu}$L of deionized water. Each coating is measured at three different positions to minimize the impact of unevenness caused by the coating. \mbox{Fig. \ref{Fig4}}(a)-(c) show the pictures of sensors with different coatings, while (d)-(e) present the corresponding CA measurement results. {It should be noted that the sensors are of the same size, as indicated by the 5 mm scale bars in \mbox{Fig. \ref{Fig4}}.}

\mbox{Table \ref{table_1}} shows measurement results of CAs and sliding angles of the three coatings. The SHP exhibits the largest CA and the smallest sliding angle, indicating the best hydrophobicity. Both RTV-SIR and RTV-SIFR coatings demonstrate similar CAs and relatively lower hydrophobicity.

\begin{table}[!t]
	\renewcommand{\arraystretch}{1.3}
	\caption{Contact Angles and Sliding Angles of Three HP Coatings}
	\centering
	\label{table_1}
	\resizebox{\columnwidth}{!}{
		\begin{tabular}{l l l l}
			\hline \\[-3mm]
			\multicolumn{1}{c}{Coating type} & \multicolumn{1}{c}{RTV-SIR} & 
            \multicolumn{1}{c}{RTV-SIFR} & 
            \multicolumn{1}{c}{SHP}  \\[1.6ex] \hline
			Contact angle ($^\circ$)  & 97.2 & 95.2 & 154.3 \\
   			Sliding angle ($^\circ$)  & 90.4 & 95.7 & 6.3 \\
			 [1.4ex]
			\hline
		\end{tabular}
	}
\end{table}

\begin{figure}[!htb]\centering
	\includegraphics[width=8.5cm]{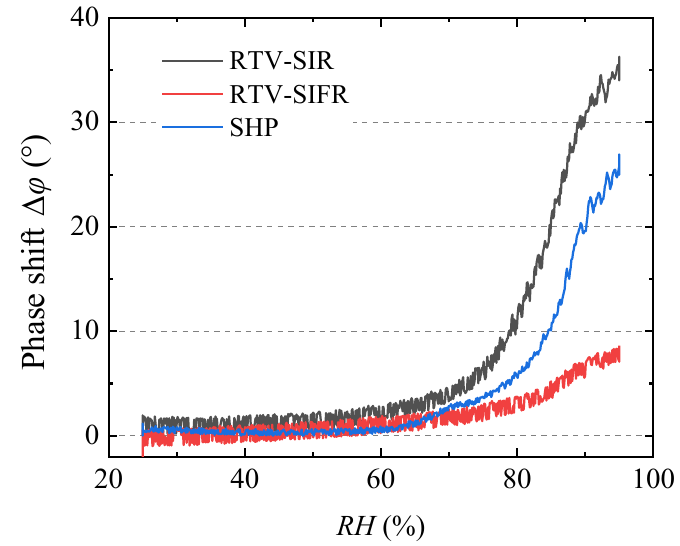}	\caption{{Measured} phase shifts of the three sensors with different HP coatings across the RH range of 25\% to 95\% at 30 $^\circ$C.}\label{Fig5}
\end{figure}

To study the humidity stability of the sensors with different HP coatings, the RH of the environmental chamber is varied from 25\% to 95\% at a rate of 5\%/h, while maintaining a temperature of 30 $^\circ$C. The {measured} phase shifts of the three sensors with different HP coatings are shown in \mbox{Fig. \ref{Fig5}}.

However, despite having the largest CA and hydrophobicity, the humidity stability of the sensor with the SHP coating is lower compared to the RTV-SIFR coating, which has the smallest CA. This indicates that the hydrophobicity of the coating is insufficient to explain the humidity stability of the sensor. Further exploration of the underlying mechanisms influencing sensor humidity stability is necessary.

\section{Influence Mechanisms of Humidity Stability}

According to the above analysis, the hydrophobicity of the coating is not the primary factor affecting humidity stability. {This section focuses on modeling and simulation of three processes: Adsorption of water vapor molecules onto the package, absorption of water vapor into the sensor shell, and adsorption of liquid water onto the package (RH $>$ 100\%).}

\begin{figure}[!htb]\centering
	\includegraphics[width=8.5cm]{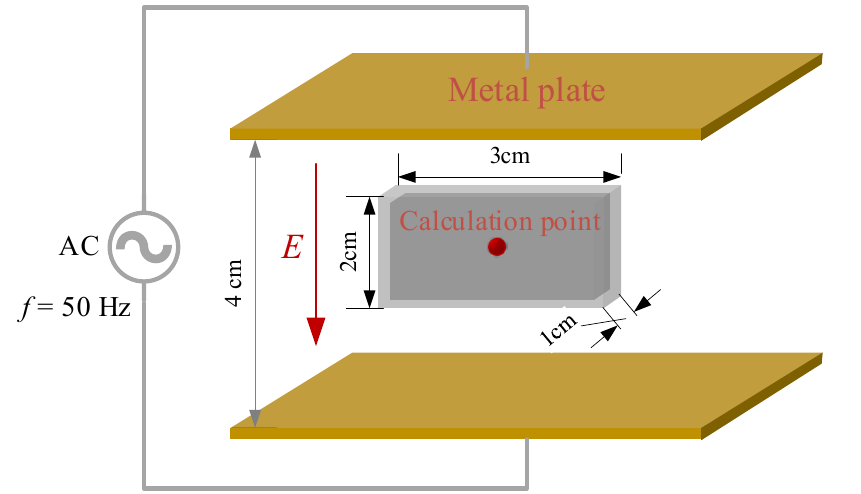}
	\caption{Simulation model of humidity stability of IOES.}\label{Fig6}
\end{figure}

The simulation layout is shown in \mbox{Fig. \ref{Fig6}}. The sensor package dimensions are 3 cm $\times$ 2 cm $\times$ 1 cm, representing a hollow structure with a thickness of 2 mm and filled with air. {A pair of 4-cm-gap} metal electrode plates are used to generate a uniform electric field. {The amplitude and frequency of the voltage applied to the upper plate are 1 V and 50 Hz,} while the lower plate is grounded. Water conductivity is set to be $5.5\times10^{-6}$ S/m, and its relative dielectric constant is 80. {The conductivity of ceramic shell is 1$\times$10$^{-14}$ S/m, and its relative dielectric constant is 4.}The HP coating of the sensor is ignored due to its thin thickness and low conductivity (about $10^{-14}$ S/m).

{Simulation models are established by the finite element method in COMSOL. The AC/DC module and transient simulation are used. The simulated sensor electric field is obtained by the probe} at the center of the package (the calculation point in \mbox{Fig. \ref{Fig6}}). The {simulated} electric field at this point is then compared with the external applied electric field {between the metal plate without the sensor inside. 

It should be noted that the change in the modal index of LN waveguide is not considered here. Due to LN’s non-hygroscopicity nature, its refractive index is considered to remain constant under different humidity levels. The refractive index variation of the surrounding air is within 1$\times$10$^{-6}$ due to different humidity levels \cite{mathar2007refractive}. Additionally, the ratio of optical mode distribution in the air cladding is very small according to \mbox{Fig. \ref{Fig1_IOES}(b).}}

\subsection{Adsorption of vapor molecules onto the {sensor shell}}

\begin{figure}[!t]\centering
	\includegraphics[width=8.5cm]{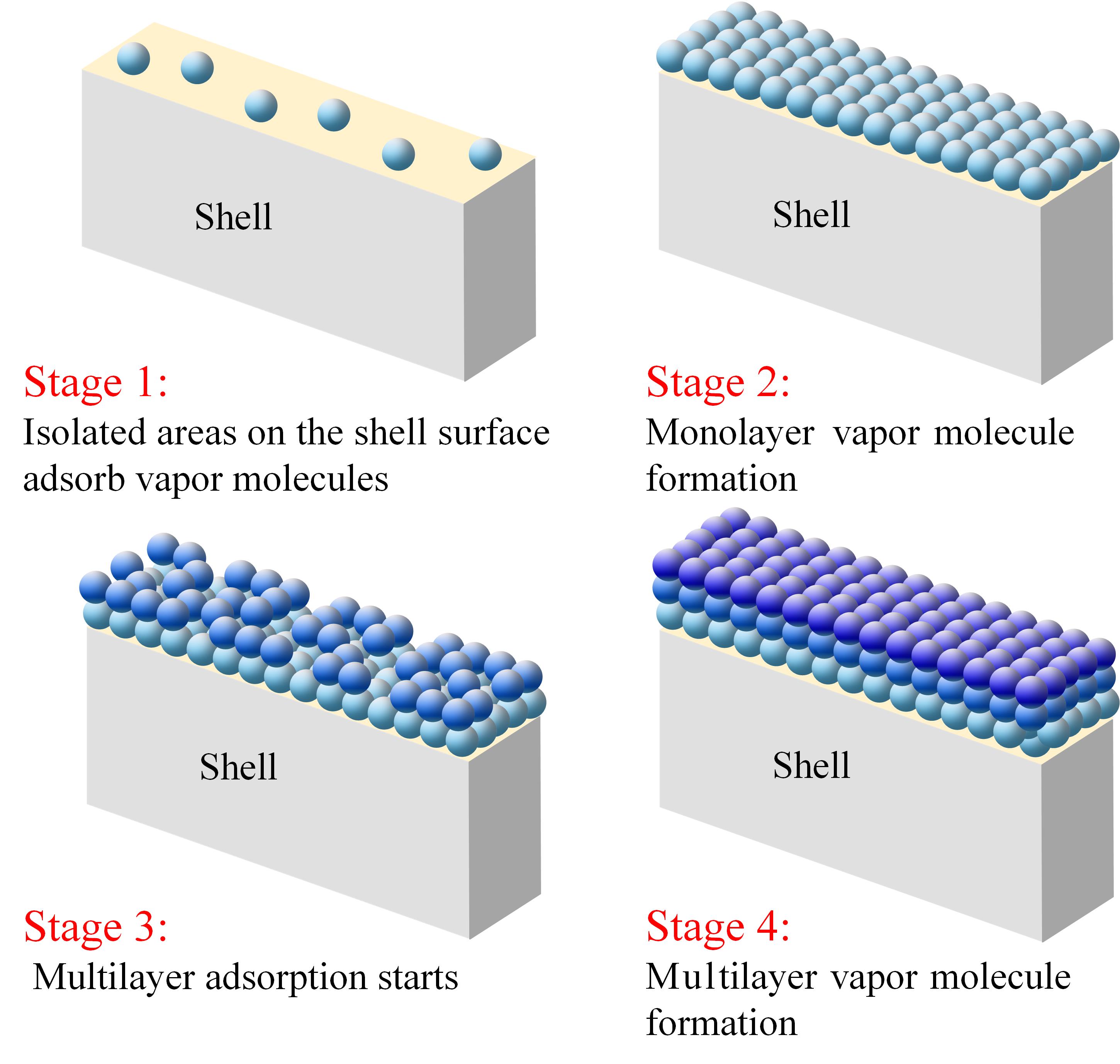}
	\caption{The physical process of vapor film formation on the shell surface of sensor.}\label{Fig7}
\end{figure}

When the RH is lower than 100\%, there is only water vapor exists in the ambient environment. Based on the multimolecular layer adsorption theory from \cite{brunauer1938adsorption}, areas without interaction on the shell surface begin adsorbing vapor molecules from the air by van der Waals forces, forming a monolayer of vapor molecule. Moreover, the van der Waals force between the monolayer and free vapor molecules, along with the long-range gravity between the shell's molecules and the vapor molecules, further promotes the adsorption of vapor molecules, resulting in the formation of a multilayer vapor molecule on the shell surface \cite{horikawa2011capillary}. The process of vapor film formation is shown in \mbox{Fig. \ref{Fig7}}.

In the process, vapor molecules are adsorbed onto the shell surface by van der Waals forces, but are also desorbed due to intermolecular collisions. When the adsorption balances the desorption, the shell's coverage is complete. Consequently, the adsorbed vapor molecules on the shell surface exist in dynamic equilibrium \cite{fagerlund1973determination}, and the film thickness is determined by substrate properties, ambient RH, and temperature \cite{ewing2006ambient}.

\begin{figure}[!htb]\centering
	\includegraphics[width=8.5cm]{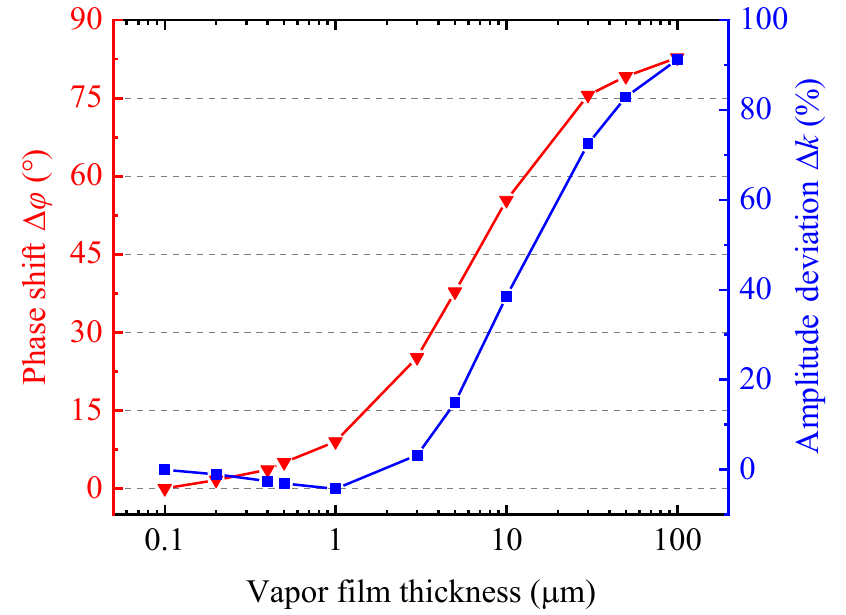}
	\caption{{Simulated }humidity stability of the sensor under different vapor film thickness.}\label{Fig8}
\end{figure}

In order to study the influence of the vapor film thickness on the humidity stability of the sensor, the sensor model shown in \mbox{Fig. \ref{Fig6}} with {different thicknesses of} vapor film on the {shell} is established. {The dielectric constant of vapor film is set to be 1}. The {simulated} relationship between the sensor output and the external applied electric field, considering different vapor film thicknesses, is shown in \mbox{Fig. \ref{Fig8}}.
The phase shift $\Delta\varphi$ and amplitude deviation $\Delta k$ of the sensor output gradually increase with the vapor film thickness. When the vapor film thickness is greater than 100 nm, $\Delta\varphi$ and $\Delta k$ are {\color{black} 83$^\circ$ and 90\%, respectively}. However, measurement results in \cite{badmann1981statistical} and \cite{zhao2015structure} show that the thickness of the vapor film on the solid surface is generally less than 10 nm under RH = 90\%. Therefore, the adsorption of water vapor molecules onto the sensor package followed by the vapor film formation has little impact on the humidity stability of the sensor.

\subsection{Absorption of vapor into the sensor shell}

\begin{figure}[!htb]\centering
	\includegraphics[width=6.5cm]{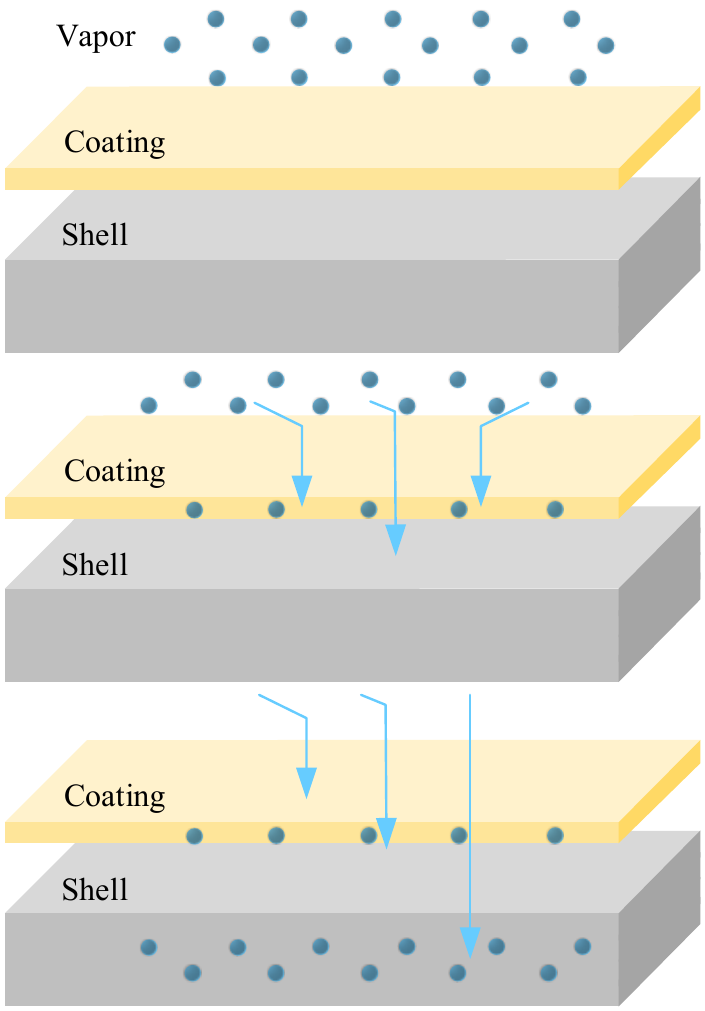}
	\caption{Model of absorption of vapor into the shell through HP coating. It is divided into three stages: vapor molecules are absorbed by the coating through capillary action; a few vapor molecules diffuse within the coating; and the majority of vapor molecules from the coatings are absorbed by the shell and diffuses into the shell.}\label{Fig9}
\end{figure}

Both the HP coating and the shell material of the sensor possess certain hygroscopicity and moisture permeability \cite{chen2014field}. The tiny pores within the HP material absorb vapor molecules from the air through capillary action. Moreover, due to the potentially higher hygroscopicity of the shell material, a substantial amount of vapor molecules within the HP material will diffuse and dissolve into the shell material \cite{harun2014drying}. The process of the shell absorbing vapor molecules is shown in \mbox{Fig. \ref{Fig9}}.

\begin{figure}[!t]\centering
	\includegraphics[width=8.5cm]{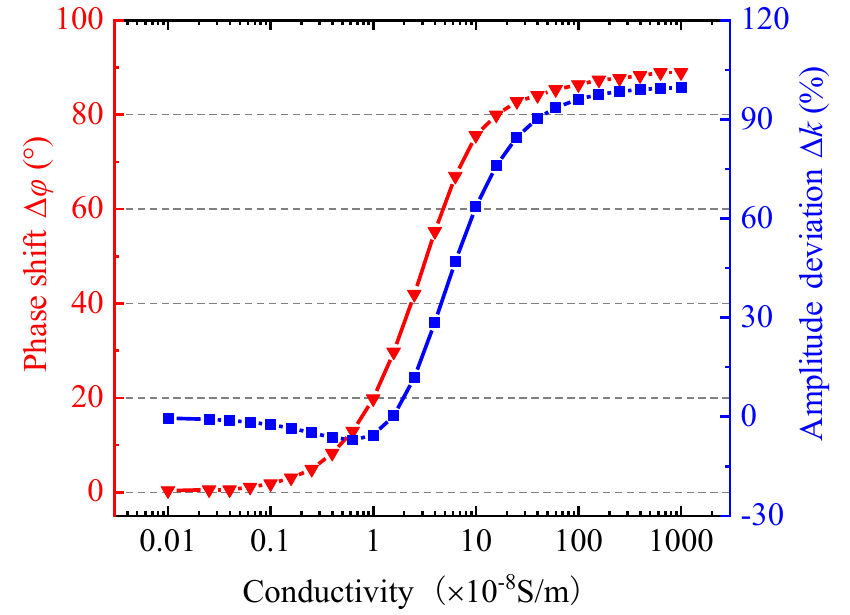}
	\caption{{Simulated} relationship between the humidity stability of the sensor and the conductivity of the shell material. The conductivity of the sensor shell indicates the water vapor absorption of the sensor.}\label{Fig10}
\end{figure}

As the RH increases, the absorption of water vapor by the shell also increases, resulting in an increase in the conductivity \cite{ogura2000effect}, \cite{awakuni1972water}. {The process is simulated by varying the conductivity of the shell material. } The {simulated} relationship between humidity stability of the sensor and conductivity of the shell is shown in \mbox{Fig. \ref{Fig10}}.

When the conductivity of shell material exceeds 1$\times$10$^{-8}$ S/m, the phase shift $\Delta \varphi$ and the amplitude deviation $\Delta k$ of the sensor increase rapidly. When the shell conductivity increases to 1$\times$10$^{-6}$ S/m, the phase shift and the amplitude deviation gradually increase to saturation, which are 90$^\circ$ and 100\%, respectively.

Therefore, the humidity stability of the coated sensor is significantly influenced by both hygroscopicity of the shell material and the moisture permeability of the HP coating. The results in \mbox{Fig. \ref{Fig5}} which shows the best humidity stability of RTV-SIFR coating may because of its low porosity and moisture permeability.

\subsection{Adsorption of liquid water onto the package}

Once the RH surpasses the saturated vapor pressure of air, {(i.e., RH $>$ 100\%)}, the water vapor molecules condense into visible liquid water and are adsorbed onto the {shell} surface.

\begin{figure}[!htb]\centering
	\includegraphics[width=8.5cm]{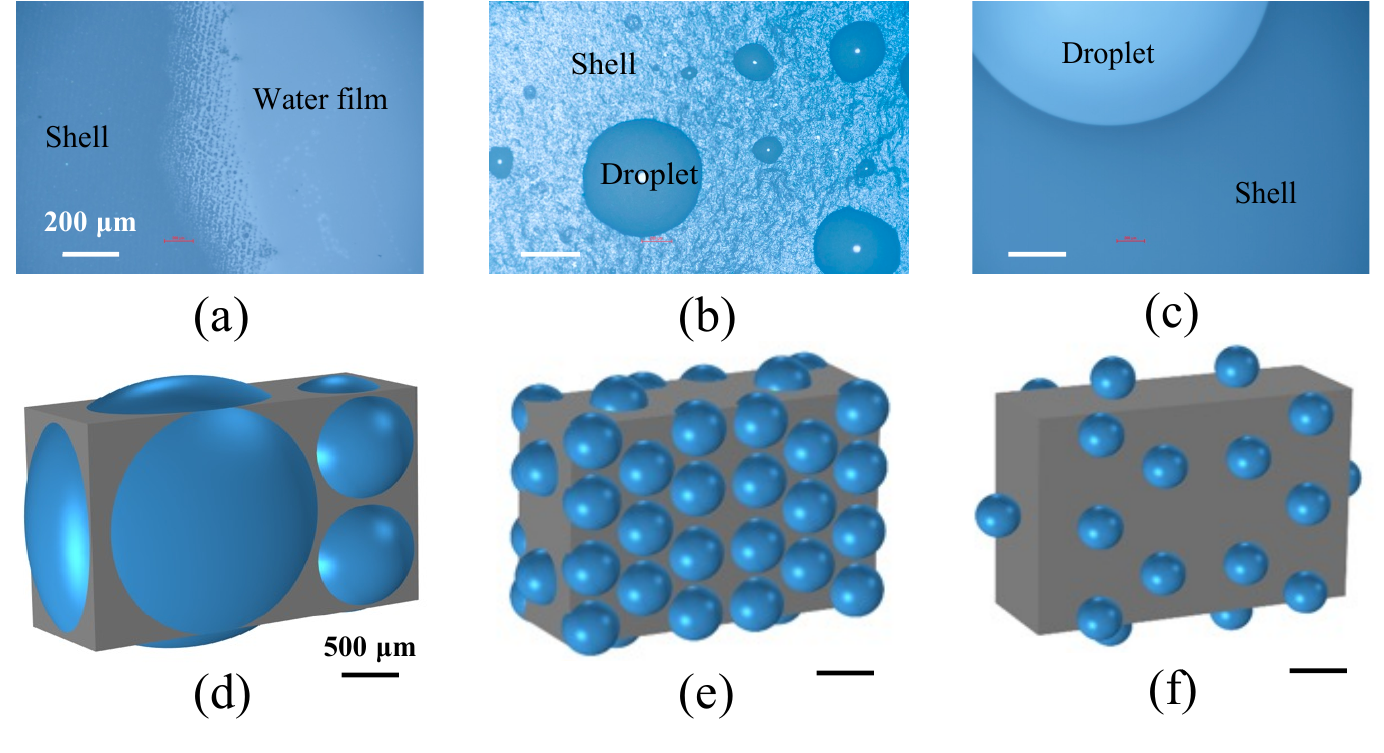}
	\caption{The morphology of the water droplets on the shell surface observed by a metallurgical microscope (a-c) and models in COMSOL (d-e). The morphology of water film on the (a) ceramic shell surface, (b) RTV-SIFR coating and (c) SHP coating. {The white and black scale bars indicate 200 $\mu$m and 500 $\mu$m in length, respectively.} The models of the shell covered with (d) continuous water film (CA $<$ 90$^\circ$), (e) separate water droplet (CA $=$ 90$^\circ$) and (f) separate water droplet (CA $=$ 150$^\circ$).}\label{Fig11}
\end{figure}

The morphology of water droplets on different HP {shell} surfaces is observed with a metallurgical microscope and modeled in COMSOL, as shown in \mbox{Fig. \ref{Fig11}}. \mbox{Fig. \ref{Fig11}}(a)-(c) represent the water droplet morphology on the surface of ceramic shell (CA $<$ 90$^\circ$), RTV-SIFR (CA = 90$^\circ$), and SHP (CA $=$ 150$^\circ$), respectively. \mbox{Fig. \ref{Fig11}}(d)-(f) show the respective water droplet models in COMSOL.

When the CA is less than 90$^\circ$, liquid water on the shell surface forms a visible water film that connects the upper and bottom surfaces of the sensor {shell}, as shown in \mbox{Fig. \ref{Fig11}}(a) and (d). Conversely, when the CA is approximately 90$^\circ$, liquid water adheres to the shell surface in the form of separate water droplets, as shown in \mbox{Fig. \ref{Fig11}}(b) and (e). As the CA increases, the contact area between the water droplets and the shell reduces, resulting in smaller water droplets remaining on the shell surface, as shown in \mbox{Fig. \ref{Fig11}}(c) and (f).

\begin{figure}[!t]\centering
	\includegraphics[width=8.5cm]{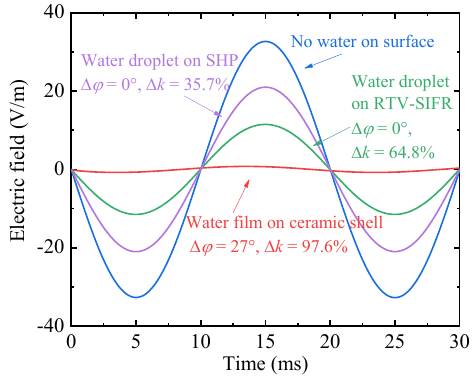}
	\caption{The {simulated} influence of the {shell} hydrophobicity on the humidity stability when there is liquid water on the surface of the sensor {shell}. The CAs of the ceramic shell,  {shell} coated with RTV-SIFR and SHP materials are $<$90$^\circ$, $=$ 90$^\circ$, and $=$ 150$^\circ$, respectively.}\label{Fig12}
\end{figure}

{The simulation models use spheres with equivalent size and conductivity to microscopic water droplets.} The {simulated electric field output} of the sensors with different degrees of shell hydrophobicity are shown in \mbox{Fig. \ref{Fig12}}.

As shown by the red curve in \mbox{Fig. \ref{Fig12}}, when the {shell} is hydrophilic, the amplitude deviation $\Delta k$ approaches 100\% and the phase shift $\Delta \varphi$ is significant. This is because the water film on the shell connects the upper and bottom surfaces of the sensor {shell}, forming a conductive surface along the applied field, which exhibits a shielding effect. When the CA of the {shell} exceeds 90$^\circ$, such as with RTV-SIFR and SHP coatings, the phase shifts $\Delta \varphi$ become nearly 0$^\circ$ since the separate water droplets on the shell disrupt the conductive path. However, $\Delta k$ is still influenced due to some degrees of the shielding effect. Comparing the RTV-SIFR and SHP coatings, the SHP coating has a smaller contact area between the shell and water droplets, resulting in lower surface conductivity of the shell. Consequently, the amplitude deviation $\Delta k$ is smaller, as shown by the green and purple curves in \mbox{Fig. \ref{Fig12}}.

Therefore, when the RH surpasses the saturated vapor pressure of air and liquid water exists in the environment, the CA and hydrophobicity of the coating influence the humidity stability of the sensor by shielding effect. A larger CA of the coating corresponds to better humidity stability.

\section{Humidity Stability Improvement and Discussion}

Based on the above analysis, a shell material with low hygroscopicity, poly-ether-ether-ketone (PEEK), is chosen as an alternative shell material besides ceramics. The material has low dielectric constant and relatively small thermal expansion coefficient. The characteristics of the two shell materials, PEEK and ceramic, are listed in \mbox{Table \ref{table_2}}.

\begin{table}[!t]
	\renewcommand{\arraystretch}{1.3}
	\caption{Characteristics of two shell materials}
	\centering
	\label{table_2}
	\resizebox{\columnwidth}{!}{
		\begin{tabular}{l l l}
			\hline \\[-3mm]
			\multicolumn{1}{c}{Material type} & \multicolumn{1}{c}{PEEK} & 
            \multicolumn{1}{c}{Ceramic}  \\[1.6ex] \hline
			Conductivity (S/cm)  & 3$\times10^{-16}$ & 8$\times10^{-11}$ \\
   			Relative dielectric constant & 3.2 & 4 \\
                Coefficient of thermal expansion (ppm/$^\circ$C) & 25-50 & 3-5 \\
                Hygroscopicity & Low & High \\
			 [1.4ex]
			\hline
		\end{tabular}
	}
\end{table}

For the ceramic materials with high hygroscopicity \cite{2008Water}, the conductivity increases by orders of magnitude with the increase of RH. Conversely, PEEK material with low hygroscopicity \cite{1999Effect}, exhibit relatively minor variations in conductivity over a wide range of RH \cite{ogura2000effect}.

\begin{figure}[!htb]\centering
	\includegraphics[width=8.5cm]{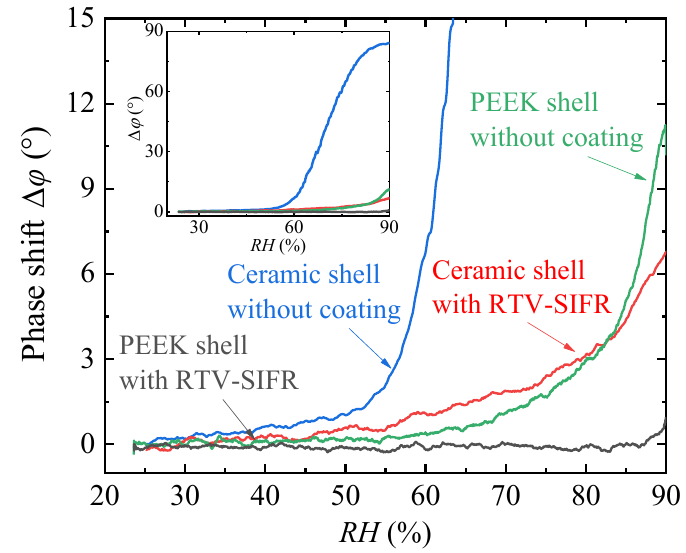}
	\caption{{Measured} phase shifts of sensors packaged with different shell materials across the RH range of 25\% to 90\% at 30$^\circ$C.}\label{Fig13}
\end{figure}

Four sensors were fabricated and tested: an uncoated sensor with ceramic shell, an RTV-SIFR coated sensor with ceramic shell, an uncoated sensor with PEEK shell and an RTV-SIFR coated sensor with PEEK shell. The experiment conditions are consistent with those depicted in \mbox{Fig. \ref{Fig5}}, and the {measured} results are shown in \mbox{Fig. \ref{Fig13}}.

For the two sensors packaged without coating (black and green curves in \mbox{Fig. \ref{Fig13}}), the humidity stability of sensor with ceramic package is found to be inferior to that of the sensor with PEEK package. The sensor with PEEK package and RTV-SIFR coating (gray curve in \mbox{Fig. \ref{Fig13}}) has the most superior humidity stability, with phase shift remaining within in 1$^\circ$ across the RH range of 25\% to 90\% and at the temperature of 30 $^\circ$C.

\begin{figure}[!t]\centering
	\includegraphics[width=8.5cm]{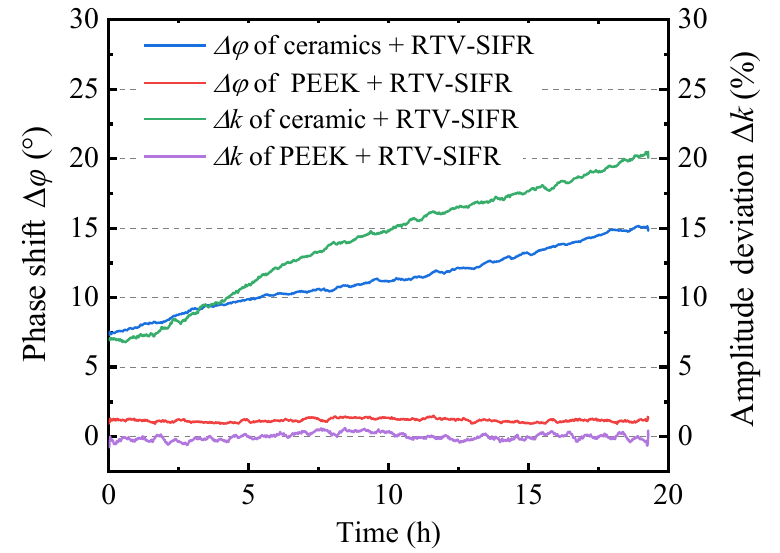}
	\caption{{Measured} long-term humidity stability of sensors with different shell materials and RTV-SIFR coating (RH = 90\%, T $=$ 30$^\circ$C).}\label{Fig14}
\end{figure}

To further investigate the long-term humidity stability of the sensor, the proposed packaging scheme is subjected to testing in a high humidity environment (RH = 90\%, T = 30 $^\circ$C) for about 20 hours. A comparison is made with the sensor packaged with ceramic shell and RTV-SIFR coating. The {measured} results are shown in \mbox{Fig. \ref{Fig14}}.

For the sensor packaged with ceramic shell and RTV-SIFR coating, the strong hygroscopicity nature of the ceramic material causes an increasing number of vapor molecules to dissolve into the shell through the HP coating, resulting in the noticeable and continuous increase of the amplitude deviation $\Delta k$ and phase shift $\Delta \varphi$ of the sensor as the duration extends. $\Delta k$ and $\Delta \varphi$ increase to 20\% and 15$^\circ$ at 20 h, respectively.

However, the combination of low hygroscopicity of PEEK material with low moisture permeability of RTV-SIFR coating results in $\Delta k$ and $\Delta \varphi$ remaining close to zero and stable over time, as shown by the red and purple curves in the \mbox{Fig. \ref{Fig14}}. The slight phase shift of about 1$^\circ$ is attributed to the HP material absorbing a minor amount of {water vapor} under the relatively high humidity (RH = 90\%).

{Humidity stability is mainly influenced by the packaging scheme rather than the change in LN waveguide. When RH increases, the sensor shell conductivity increases because of water vapor absorption. The IOES shell has its resistance and capacitance, which form a parallel resistor-capacitor (RC) network with the LN waveguide's resistance and capacitance. An increase in sensor shell conductivity results in a decreased resistance in the RC network, which further results in a decreased output amplitude and an increased phase shift in the sensor. The humidity-stable packaging scheme proposed here uses PEEK shell, which has moisture permeability, with room temperature vulcanized fluorinated silicone rubber (RT-SIFR) coating, which has low hygroscopicity.}

Considering the outdoor application environment for sensors, both high RH and the presence of liquid water are likely to occur. Therefore, it is proposed that the RTV-SIFR and SHP coatings could be used together, with the SHP coating as the outmost layer to prevent liquid water. The packaging scheme has the potential to improve the humidity stability of other outdoor integrated sensors as well.

\section{Conclusion}

The paper studies the influence mechanisms on the humidity stability of the integrated optical electric field sensor. It reveals that the hydrophobicity of the sensor {shell} only affects the humidity stability when there is liquid water on the {shell} surface. In environments where only water vapor is present, the hydrophobicity of the {shell} and the thickness of vapor film have no obvious impact on the humidity stability. Instead, the hygroscopicity of the shell and the moisture permeability of the coating play vital roles in the humidity stability. This clarification corrects the misunderstanding that the hydrophobicity of sensor {shell} alone determines the humidity stability of the sensor.

An optimal packaging scheme is proposed, featuring a combination of the low hygroscopicity shell material PEEK and the low moisture permeability HP coating RTV-SIFR. The sensor with the proposed packaging scheme demonstrates phase shifts consistently within in 1$^\circ$ across the RH range of 25\% to 90\% at 30 $^\circ$C. The long-term humidity stability shows {\color{black}an improvement in amplitude deviation of the electric field measurement from 20\% to nearly zero} after 20 hours in a high humidity environment (RH = 90\%, T = 30 $^\circ$C). A double-layer coating with SHP besides the RTV-SIFR material could be used to resist liquid water as well. The proposed packaging scheme could also be extended to improve the humidity stability of other outdoor integrated sensors.


\bibliographystyle{IEEEtran}
\bibliography{IEEEabrv, ref.bib}\ 

\end{document}